\documentclass[structabstract]{aa}  
\usepackage{graphicx}
\usepackage{txfonts}
\begin{document}

\title{A double detached shell around a post-Red Supergiant: \\
IRAS~17163-3907, the 
Fried Egg nebula \thanks{Based on observations made with the Very Large Telescope
at Paranal Observatory under program 081.D-0130(A).}\fnmsep\thanks{Based on observations made with the Mercator Telescope,
 operated on the island of La Palma by the Flemish Community, at the Spanish Observatorio del Roque de los Muchachos of the Instituto de Astrofísica de Canarias.}}


   \author{E. Lagadec
          \inst{1}, A.A.
          Zijlstra\inst{2}, R.D. Oudmaijer\inst{3} , T. Verhoelst\inst{4}, N.L.J. Cox\inst{4}, R. Szczerba\inst{5},  D. M\'ekarnia\inst{6} and  H. van Winckel\inst{4}}

   \institute{European Southern Observatory, Karl Schwarzschildstrasse 2, Garching 85748, Germany \and Jodrell Bank Center For Astrophysics, The University of
   Manchester, School of Physics \& Astronomy, Manchester M13 9PL, 
   UK \and School of Physics and Astronomy, University of Leeds, Leeds LS2 9JT, UK \and Instituut voor Sterrenkunde, K.U. Leuven, Celestijnenlaan,
   200D, 3001, Leuven, Belgium \and N. Copernicus Astronomical Center,
   Rabia\'nska 8, 87-100 Tor\`un, Poland  \and  Lab. H. Fizeau, CNRS UMR 6525, 
   Univ. de Nice-Sophia Antipolis, Observatoire de la C\^ote d'Azur,
   06108 Nice, France }
         
\titlerunning{A double detached shell around the Fried Egg nebula}
\authorrunning{Lagadec et al.}
   \date{Received ; accepted}

 
  \abstract
   {We performed a  mid-infrared imaging survey of evolved stars in order to study the dust distribution in circumstellar envelopes around these objects 
and to better understand the mass-loss mechanism responsible for the formation of these envelopes. During this survey, we resolved for the first time the 
circumstellar environment of IRAS~17163-3907 (hereinafter IRAS17163), which is one of the brightest objects in the mid-infrared sky, but is surprisingly not well studied.}
   {Our aim is to determine the evolutionary status of IRAS~17163 and  study its circumstellar environment in order to understand its mass-loss history.}
   {We obtained diffraction-limited images of IRAS 17163 in the mid-infrared using VISIR
   on the VLT. Optical spectra of the object allowed us to determine its spectral type and estimate its distance via the presence of diffuse interstellar bands.}
   {We show that IRAS~17163 is a Post-Red Supergiant, possibly belonging to the rare class of Yellow Hypergiants, and is very similar
 to the well studied object IRC~+10420. Our mid-infrared images of IRAS 17163 are the first direct images of this bright
   mid-infrared source. These images clearly show the presence of a double dusty detached
   shell around the central star, due to successive ejections of material  with a timescale of the order of 400 years and a total circumstellar mass larger 
than  4~M$_{\odot}$. This indicates that non quiescent
mass-loss occurs during this phase of stellar evolution.

   }
   {}

\maketitle

\section{Introduction}
After exhausting the hydrogen in their core at the end of the main sequence,
stars with initial masses in the range 8-40 M$_{\odot}$ become large and cool 
Red Supergiants. When they leave the Red Supergiant branch, those 
stars evolve via the Yellow Hypergiant phase, followed by the Luminous
Blue Variable phase (LBV), to finally become  Wolf-Rayet stars (Oudmaijer et al., 2009).The stellar winds 
associated with 
these phases, followed by a final supernova
explosion, are essential for the chemical enrichment of galaxies and provide kinetic energy that can 
trigger star formation. These stellar winds are likely due to a combination of pulsation
and radiation pressure on dust. This leads to the formation of a dusty circumstellar 
envelope which makes  post-main sequence objects bright infrared sources.

As part of a mid-infrared imaging survey of evolved stars (Lagadec et al. 2011), we 
observed the very bright mid-infrared source 
(with a 12~$\mu$m flux F$_{12}$=1243 Jy)
IRAS 17163-3907 (hereinafter IRAS17163). 

IRAS 17163 (a.k.a. Hen 1379 ) was discovered by Henize in 1976 during a survey 
of emission line stars in the southern sky.
From their  2.2~$\mu$m mapping (Valinhos 2.2 micron survey), Epchtein et al. (1987)
classified this object as a PPN candidate. 
Optical spectra of this object allowed us to show that the distance to this object is four times larger 
than previously assumed, and thus too bright to be a post-AGB star.

We report here  the direct detection of a double detached shell around the central star 
of this post-Red Supergiant object.


\begin{figure*}
\includegraphics[width=18cm]{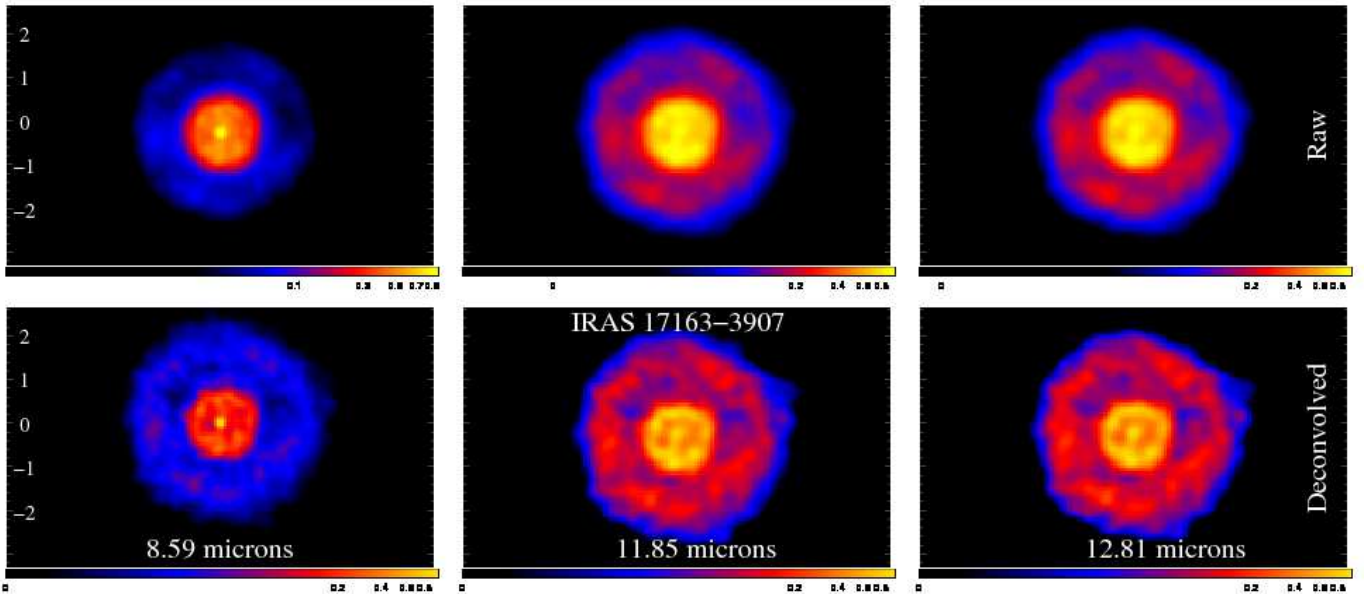}
\caption{\label{visir_17163} VISIR/VLT mid-infrared images of IRAS 17163 at different wavelengths. The raw and deconvolved images are displayed for each filter.
 North is up and East left. The spatial scale is in arcsec and the intensity scale indicates the relative intensity. We clearly see
the central star and two dusty detached shells.}
\end{figure*}
\vspace{-0.5cm}
\section{Observations and data reduction}
We observed IRAS17163 in the mid-infrared with VISIR on the VLT (Lagage et al., 2004) with 3 filters: PAH1 (8.59~$\mu$m, half bandwidth 0.42~$\mu$m ), 
SiC (11.85~$\mu$m, 2.34~$\mu$m) and NeII (12.81~$\mu$m, 0.21~$\mu$m).
We used the imager in burst mode, using a pixel scale of 0.075~arcsec and a field of view of 
19.2 $\times$ 19.2 arcsec. With the burst mode, all the  chopping and nodding images 
are recorded,  allowing the reconstruction of quality-enhanced images using shift and add 
techniques. We used the standard chopping/nodding technique to remove the background. 
We shifted and added the images using a maximum of correlation algorithm, after removing the
bad images.
Our observations were obtained during one of the driest nights ever in
Paranal (0.43~mm of precipitable water vapor in the atmosphere). The quality of mid-infared
data being mostly affected by humidity, we obtained diffraction-limited
images with  a great stability (seeing+sky background)  throughout the whole observating
run . The images were deconvolved using the standard star HD 163376
 as a measure of the Point Spread Function (PSF,  observed before and after the science target with the same integration time of 30 seconds) and a maximum of likelihood algorithm with 
 50 iterations.
Fig.\,\ref{visir_17163} shows the  raw and deconvolved images in the three filters.

We obtained an optical spectrum of IRAS 17163 using the
HERMES instrument (High Efficiency and Resolution Mercator Echelle
Spectrograph; Raskin et al., 2011) on the Mercator telescope (1.2~\,m) in
La Palma.  HERMES is a high-resolution fiber-fed echelle spectrograph
 offering a spectral resolution R $\sim$ 80\,000 and a 2.54~arcsec field of view. We obtained 4 exposures 
of 1200\,seconds each, under mediocre seeing conditions, resulting in a SNR
 of 20 at 6000\,\AA.  The pipeline-reduced spectrum stretches from 3770 
to 9900\,\AA, but due to the red nature of the source, it only appears above the noise from 5300\,\AA\  onwards.

\vspace{-0.5cm}

\section{A new Post-Red Supergiant star}
\label{dist}

Diffuse interstellar Bands (DIBs), due to absorption by the interstellar medium, are seen  
in our optical spectrum of IRAS~17163. This allows us to estimate the distance to the observed
object.

\begin{figure}[h]
\includegraphics[width=8cm]{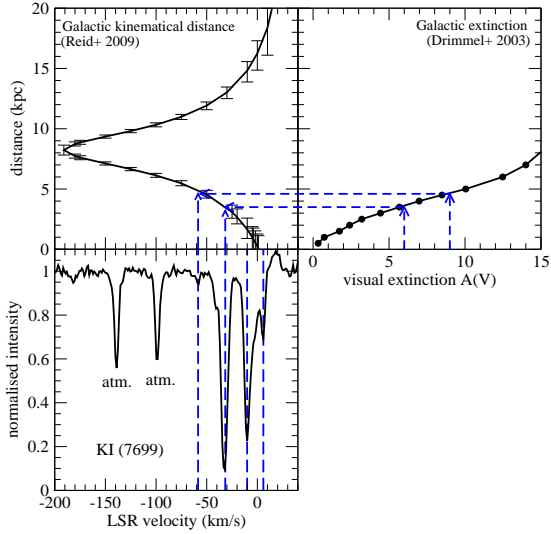}
\caption{\label{vel_dis}The \ion{K}{i} doublet velocity profile (lower left) of IRAS~17163, together with the Galactic kinematical distance estimate (upper left)  from Reid et al (2009) and the model 
of Galactic extinction (upper right) from Drimmel et al. (2003). This indicate that the distance of IRAS~17163 is between 3.6 and 4.7 kpc.}
\end{figure}

We adopted the relation defined by Reid et al. (2009) to calculate the kinematic distances 
for
interstellar clouds observed in the line-of-sight toward IRAS~17163.
The \ion{K}{i} doublet velocity profile is plotted in Fig.~\ref{vel_dis}.
Three main interstellar velocity components are observed at local
standard of rest (LSR) velocities of -31.7, -10.1, and 5.9~km\,s$^{-1}$. A
possible weak component at -58.5~km\,s$^{-1}$ is seen in both KI profiles.
In addition, we observe narrow (FWHM $\sim$ 0.15~\AA) interstellar C$_2$
absorption lines  at a LSR velocity of -32.3~km\,s$^{-1}$.

For the interstellar cloud at $\sim$-32~km\,s$^{-1}$ we obtain the strongest
constraints on the kinematic distance.
We find a lower limit of 3.6~kpc for this interstellar cloud, which is
also a lower limit of the distance to IRAS 17163.

If the Galactic rotation model is applied to the  component at
5.9~km\,s$^{-1}$ this would give a {\it lower} limit of
17.4$^{+1.84}_{-1.33}$~kpc to the star. This, however, is incompatible with
the inferred total line-of-sight reddening.
We consider it most likely that this diffuse cloudlet has a peculiar
velocity with respect to the rotational/radial velocity of the Galactic
disk in this direction. The weak \ion{K}{i} cloud at -58.5~km\,s$^{-1}$ would
correspond to a minimum distance of 4.7~kpc.

The \ion{K}{i} velocity absorption profile is shown together with both
the Galactic radial velocity (in LSR) and the visual extinction $A_V$ as
a function of distance.

From the measured equivalent widths for the 5780, 5797,
6379, and 6613~\AA\ DIBs we infer a reddening, E($B-V$), between 1.6 and
2.8~mag  (Luna et al., 2008), with a mean of 2.1~mag (Table~\ref{table:DIB}). For typical ISM dust, the total-to-selective visual
extinction is $R_V$ = 3.1 (Fitzpatrick \& Massa 2009) and leads to a visual extinction of  6.4 magnitudes. The maximum value
for the reddening corresponds to $A_V \sim 8.7$~mag. In Fig.~\ref{vel_dis} we
plot the visual extinction versus distance extracted from the 3D Galactic
dust extinction map constructed by Drimmel et al. (2003). We also
indicate the distances derived from interstellar visual extinction values
of 6 and 9~mag towards IRAS~17163. Although the Galactic dust extinction model may
not be entirely accurate for sightlines within $\sim$ 20 degrees toward  the Galactic center, in
this case the different distance determinations are fully consistent
with each other.

\begin{table}
\caption{Measured DIB strengths (in \AA, with 10\% uncertainties)
towards IRAS 17163.
The 8621~\AA\ DIB appears blended with a \ion{H}{i} emission line and
was thus rejected for the calculation of E(B-V). EW/E(B-V) are taken from Luna et al.(2008).}
\label{table:DIB}
\begin{tabular}{llll}\hline\hline
DIB       &    EW (\AA)     &        EW/E(B-V)    &  E(B-V) for
IRAS 17163 \\ \hline
5780    &    0.91              &               0.46               &    
1.98    \\
5797    &    0.48               &             0.17                &    
2.82   \\
6379    &    0.14               &            0.088               &    
1.59  \\
6613    &    0.404             &              0.21              &       
1.92 \\
8621    &    0.497             &            0.37                &      
 1.34   \\     
\hline
\end{tabular}
\end{table}

In conclusion, the strong interstellar component at -32\,km\,s$^{-1}$  and the observed
visual extinction, $A_V$\,$ \geq$\,$ 6$\,mag  suggest a distance
of 3.6\,kpc to this IS cloud, and  provide a lower limit to the
distance of IRAS\,17163.
The maximum visual extinction ($\sim$\,$9$\,mag)  consistent with the measured DIBs strengths
  implies an upper limit to the distance of $\sim$\,4.7\,kpc.
Therefore, the distance to IRAS\,17163 is between 3.6 and 4.7\,kpc.

This is four times further than what was previously assumed by Le Bertre et al. (1989), who classified
this star as a post-AGB star. The luminosity of
IRAS~17163 ( 5$\times 10^5$ L$_{\odot}$, see Sect.\,\ref{yellow}) is thus well above the maximum luminosity of a post-AGB star ($\sim 10^4$L$_{\odot}$; Schoenberner et al., 1983).
IRAS~17163 is  clearly not a post-AGB star, but very likely a post-Red Supergiant.

\vspace{-0.4cm}
\section{The double dust shell}

  \begin{figure}[h]
\includegraphics[width=8cm]{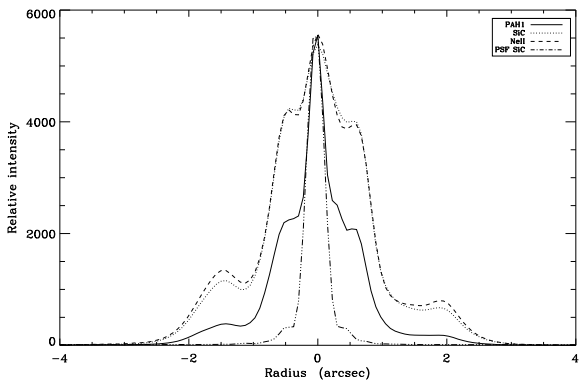}
\caption{\label{prof_dec} Radial profile of the non deconvolved images in the three VISIR  filters, showing the presence of the star and 
a double detached shell. The PSF is also shown for reference.}
\end{figure}  

Le Bertre et al. (1989) obtained observations of IRAS 17163
and concluded  that IRAS 17163 was located at  1~kpc from the Sun, with a luminosity  well below the maximum for post-AGB stars.
 The high polarization ($p\sim 12\%$)
measured  in V, R and I seems to  indicate that the circumstellar envelope  is highly aspherical.
IRAS 17163 is unresolved in the optical images obtained with the ESO 1.5m danish
telescope at  la Silla (Chile),
 (Le Bertre et al., 1989), while infrared speckle interferometric observations
 indicate an angular dimension of  1.11$\pm 0.23$" in L band (3.6~$\mu$m) 
 (Starck et al., 1994). 
The high resolution H$_{\alpha}$ profile of IRAS~17163 indicate a  wind
velocity of $\sim$63~km\,s$^{-1}$.

IRAS~17163 was observed by the HST in the optical (Si\'odmiak et al., 2008), but classified as not resolved.
 The integration time was short (45s) and after inspection of the HST image, it appears that the dimension of 
the detached shells correspond to the Airy pattern of the HST at the observed wavelengths, which is 
why the authors did not see the detached shells (Si\'odmiak, private communication).
The images of IRAS~17163 we obtained with VISIR are displayed Fig. \ref{visir_17163}. 
The object is 
clearly resolved in all the filters,
with a diameter of $\sim$5". These are the first direct images of the 
circumstellar material 
around this very bright infrared source.
The mid-infrared morphology of IRAS~17163 
is similar to a fried egg, and 
we dubbed it the "Fried Egg" nebula.
The object is circular at large scale and the central star can be observed 
in all the filters. 

\begin{figure}[h!!!!]
\includegraphics[width=9cm]{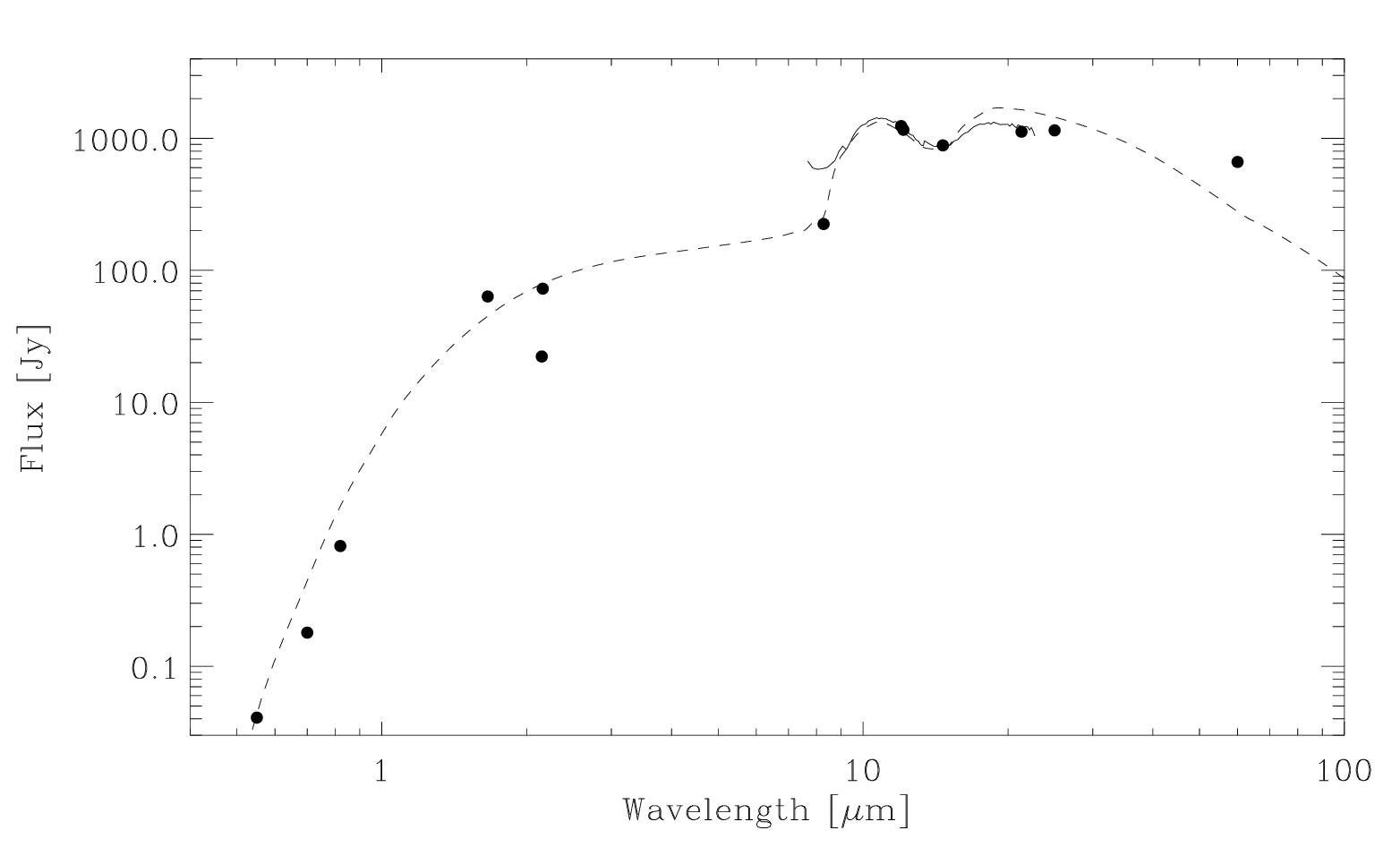}
\caption{\label{SED} Spectral energy distribution of IRAS 17163, the data points are from Szczerba et al., 2007. The dashed line represents our best radiative transfer model fit to the data.}
\end{figure}

The most striking aspect of these images is the presence of a double detached 
shell around the central star.
This is more clearly seen in Fig.\ref{prof_dec} which shows a cut along the 
North-South direction in the 
three deconvolved images.
We can clearly see two shells at $\sim$0.6 and 1.5" from the central star.
 The images show that those shells are quite round. The shells
 however show some clumpy structures
 that are apparent in all the images and are thus not deconvolution artifacts.
 While on the raw images, the
 shells seems to be very smooth,  the deconvolved images clearly show that these 
 shells are made of a large
 number of clumps. We can also  see a large hole in the second dusty shell along 
 a P.A.of  $\sim$45$^o$.
An extended structure, which seems to link the central star and the first shell 
is also seen along a P.A. $\sim$170$^o$.
There are some signs for a departure from spherical symmetry in the core of the nebula.

Considering a distance to the object of 4~kpc, the two shells have a radius of 
2400 and 6000 A.U. respectively.
Assuming a distance of 4~kpc and an expansion velocity of 40~km\,s$^{-1}$, this leads
to an estimate of $\sim$435~years for the time
interval between the two shells.

To estimate the amount of dust present in the observed shells, we developed a radiative transfer 
model. We used the code described by Szczerba et al. (1997).  The aim of our model is to fit both the SED and the intensity profile in our three VISIR images
to get an estimate of the mass of dust in the envelope. The IRAS spectrum of IRAS 17163 
 shows the presence
of silicate features,   indicating that  O-rich dust is present in
the circumstellar envelope. Assuming a distance of 4~kpc, a star effective temperature of 8000~K and that the dust is made of circumstellar
silicates with a standard MRN  size distribution (Mathis et al., 1997), we obtained the best fit to the SED and radial dust distribution with a total
dust mass of 0.04~M$_{\odot}$,  and  a mass-loss rate varying with r$^{-1}$ between the shell, as the SED could not be fitted without the presence of dust in between the shells. 
  Assuming a lower limit of 100 for the gas-to-dust mass ratio, this leads to a total gas mass of 4~M$_{\odot}$ in the ejecta and an average mass-loss rate 
of 10$^{-4}$M$_{\odot}$yr$^{-1}$ and 2-3$\times$10$^{-3}$M$_{\odot}$yr$^{-1}$  associated with the formation of the inner and outer shells respectively .
Some dust is present in the inter-shell, and the presence of hot dust ($\sim$1500~K) close to the central has to be assumed to
 reproduce the flux of the central unresolved source. In the first shell  the mean dust temperature ranges from 230 to 180 K, while in the outer shell it ranges from 150 to about 110 K. 
Assuming a width of 0.5 arcsec for each shell, we found that 0.002 and 0.005~M$_{\odot}$ of dust are present in the inner  and outer shells respectively.

\vspace{-0.5cm}
\section{Discussion}
\subsection{A Yellow Hypergiant}
\label{yellow}
\begin{figure}[h]
\includegraphics[width=9cm]{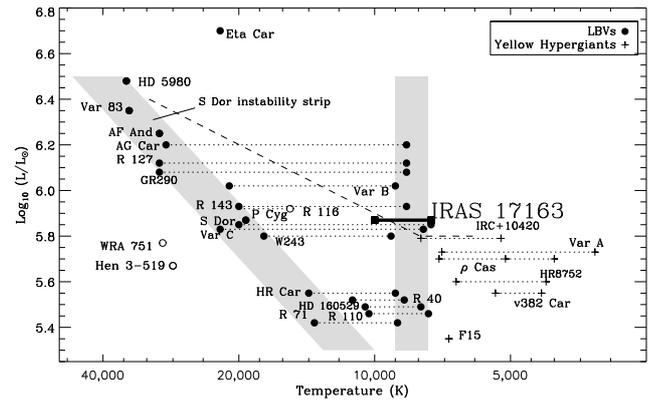}
\caption{\label{hr_diag} Temperature-luminosity diagram of post-Red Supergiants, showing that IRAS~17163 is likely to be a Yellow Hypergiant.  The grey bands on the left
and right of the diagram represent respectively the regimes in the HR diagram which the LBVs occupy when they are in
 the quiescent phase (left) and
 the active phase (right) at maximum visual brightness (Smith et al. 2004.)}
\end{figure}

The optical spectrum of IRAS~17163 shows the presence of the 7450~$\AA$ NI line and the absence of Helium lines, which indicates that the central star is either
 a late B or early A central star (Gray \& Corbally, 2009), with an effective temperature in the range
7500-10000 K. If we assume, as estimated by the DIBs and the \ion{K}{i} radial velocity we observed (Sect.\ref{dist}), a distance of
4 kpc for the object, its luminosity (measured by integrating the SED,  (Fig. \ref{SED})) is   $\sim$5$\times 10^5$ L$_{\odot}$. If one  compares the location of IRAS~17163
 on a  temperature-luminosity diagram (Fig. \ref{hr_diag}), 
as the one presented by Oudmaijer et al.(2009), we can see that its properties are very similar to the brightest Yellow Hypergiants
 like the famous IRC+10420, with its temperature and luminosity just below the Luminous Blue Variables (LBVs).

\subsection{Mass-loss variations}
 Yellow Hypergiants are not in a quiet evolutionary phase (Oudmaijer, 2009)
and display  episodic  ejections,  moving  alternatively from the blue  to the red on an HR diagram. 
The two detached shells we observe are  likely due to such  events.
 HST imaging of the Yellow Hypergiant IRC~+10420  has shown the
presence of partial shell and knot structures, which represent episodic
ejection events (Tiffany et al. 2010, and references therein).
 We observed two shells due to phases of enhanced mass-loss separated by $\sim$400~years.
The central stellar source in the 10-micron image cannot be fitted with a stellar SED, which could be due to  hot dust close to the star, indicating
 the presence of a third, unresolved  shell next to the central star.
The presence of clumps of dust close to the central star could explain the polarization observed for IRAS~17163. The unusual high linear polarization could also be partly due 
to alignment of interstellar grains in this line of sight. Polarization measurements of field stars near IRAS~17163 would be needed to resolve this issue.
It is possible that many other shells are present further away from the central star. Such shells are not observed  because of the small field of view of VISIR,
the fact that these shells would be less dense and thus  weaker and/or the fact that dust further away would be colder and thus emit at longer wavelengths.  Such shells would
 explain the discrepancy between our radiative transfer model and the observed flux at 60 microns. Millimeter observations of
the Fried Egg should allow the detection of such shells. 

\vspace{-0.5cm}
\section{Conclusions}
We presented the first resolved images of IRAS 17163. The morphology
we observed led us to dub this object the ``Fried Egg'' nebula. We resolved 
two large concentric spherical dusty shells around the central star, itself embedded in dust.
More than 4~M$_{\odot}$ of gas and dust have already been ejected by the star.
 IRAS 17163 was previously classified as a post-AGB star, but the
presence of DIBs in the optical spectra we obtained led us to the conclusion 
that the object is four times further than previously suggested and is thus a
post red supergiant. The location of the Fried Egg on an HR diagram and the
remarkable similarity between its optical spectra and the one of the prototype
 Yellow Hypergiant IRC~+10420 indicate that the Fried Egg is probably a Yellow Hypergiant.
 The presence of detached shells around IRAS~17163 is a confirmation that
non quiescent mass-loss occurs during this phase of stellar evolution.




\begin{acknowledgements} 
The authors would like to thank the anonymous referee for providing 
us with constructive comments and suggestions.  
\end{acknowledgements}

\end{document}